\documentclass[10pt,aps,pre,twocolumn,notitlepage,superscriptaddress,floatfix]{revtex4-1}
\usepackage{graphicx,epsf,amsmath,amssymb,wasysym,verbatim,color,cleveref}

\begin{document}
\title{Double transitions and hysteresis in heterogeneous contagion processes}
\author{Joongjae Kook}
\affiliation{Department of Physics, Chungbuk National University, Cheongju, Chungbuk 28644, Korea}
\author{Jeehye Choi}
\affiliation{Research Institute for Nanoscale Science and Technology, Chungbuk National University, 
	Cheongju, Chungbuk 28644, Korea}
\author{Byungjoon Min}
\email{bmin@cbnu.ac.kr}
\affiliation{Department of Physics, Chungbuk National University, Cheongju, Chungbuk 28644, Korea}
\affiliation{Research Institute for Nanoscale Science and Technology, Chungbuk National University, 
	Cheongju, Chungbuk 28644, Korea}
\date{\today}

\begin{abstract}
In many real-world contagion phenomena, the number of contacts to spreading 
entities for adoption varies for different individuals. Therefore, we study a model 
of contagion dynamics with heterogeneous adoption thresholds. We derive mean-field 
equations for the fraction of adopted nodes and obtain phase diagrams in terms of 
the transmission probability and fraction of nodes requiring multiple contacts for 
adoption. We find a double phase transition exhibiting a continuous transition and 
a subsequent discontinuous jump in the fraction of adopted nodes because of the 
heterogeneity in adoption thresholds. Additionally, we observe hysteresis curves 
in the fraction of adopted nodes owing to adopted nodes in the densely connected 
core in a network. 
\end{abstract}
\maketitle

\section{Introduction}

The spread of information, fads, rumors, innovations, or diseases has significantly 
increased globally because of the development of communication and mobility 
technologies~\cite{goffman1964,daley1964,granovetter1978,may1991,
pastor2001,watts2002,pastor2015}. To understand and control the 
spreading phenomena, quantitative modeling of contagion processes 
is crucial~\cite{pastor2015,centola2018}. Contagion processes can be divided 
into two classes based on the type of contacts with active neighbors: simple 
and complex contagions~\cite{kermack1927,harris1974,may1991,granovetter1978,
centola2007,centola2010}. Simple contagion is a contagion process with 
independent interactions between the inactive and the active \cite{may1991}. 
Compartmental epidemic models such as the susceptible-infected-recovered~\cite{may1991,
kermack1927,newman2002} and susceptible-infected-susceptible models~\cite{may1991,pastor2001} 
are examples of the simple contagion model. In contrast, complex contagion consists 
of collective interactions among active neighbors~\cite{watts2002,centola2007,chae2015,dual2018}. 
Compared to the simple contagion, complex contagion processes are controlled by group 
interaction; that is, the probability of adoption strongly depends on the 
number of contacts with active neighbors. Multiple contacts to spreading entities are 
required for adoption in the complex contagion. Various social and biological spreading 
models such as threshold model~\cite{granovetter1978,watts2002,gleeson2007,lee2014}, 
generalized epidemic model~\cite{janssen2004,choi2018,baek2019}, diffusion 
percolation \cite{adler1988}, and bootstrap percolation~\cite{chalupa1979,adler1991,baxter2010} 
are examples of the complex contagion model.

Recently, there have been several attempts for unifying simple and complex contagion 
processes \cite{dodds2004,cellai2011,baxter2011,karampourniotis2015,czaplicka2016,
wang2016,min2018}. Empirical data of the spread of social behavior in a social networking 
services (SNS) indicate that adoption thresholds for different individuals are 
heterogeneous, depending on the characteristics of agents \cite{dow2013,state2015}.
A unified contagion model incorporates different number of contacts to a spreading 
entity required for adoption, similar to the heterogeneity of adoptability observed 
in empirical data~\cite{min2018}. In particular, while some agents change their state 
in contagion processes immediately after the first contact to new information 
following simple contagion, others require multiple contacts for adoption following 
complex contagion \cite{state2015}. According to the observations, real-world contagion 
phenomena are neither simple nor complex contagions, but they could be a mixture of the 
two, with different adoption thresholds for different agents. Because the characteristics 
of individuals significantly vary in reality, the heterogeneity of adoption threshold 
can be widespread for many contagion phenomena. In addition, such a unified contagion model 
can be useful for analyzing the spread of epidemics of multiple interacting pathogens 
because interacting epidemics is indistinguishable from social complex 
contagion~\cite{janssen2016,min2020,hebert2020}. Therefore, generalized contagion processes 
integrating simple and complex contagions are of importance.

Here, we study a model of contagion dynamics on random networks with heterogeneous 
adoption thresholds for different agents. 
Although some previous works have also considered heterogeneous adoption thresholds~
\cite{dodds2004,cellai2011,baxter2011,karampourniotis2015,czaplicka2016,wang2016,min2018},
most of them have assumed that adopted individuals acquired permanent adoption.
In reality, many adopted individuals can lose their adopted state 
and return to the pool of susceptible individuals. In our model, we have 
added a transition from the adopted state back to susceptible state 
in contrary to previous studies~\cite{pastor2001,majdandzic2014}. 
The process is implemented autonomously for each agent, similar to the recovery 
process in the classical susceptible-infected-susceptible model~\cite{may1991,pastor2001}. 
In our study, we have added both heterogeneous adoption thresholds for each individual
and also recovery process. Incorporating the heterogeneity and recovery process, we 
find a double phase transition with an intermediate phase in which nodes following 
simple contagion are adopted but nodes with complex contagion remain susceptible. 
In addition, we find hysteresis curves in the fraction of adopted nodes with respect 
to the contact probability. Our study provides the simple mechanism of 
hysteresis~\cite{majdandzic2014,min2014,chen2017} 
and multiple phase transition~\cite{colomer2014,allard2017,min2018}
in contagion processes on complex networks.

\section{Model}

\begin{figure}
\includegraphics[width=\linewidth]{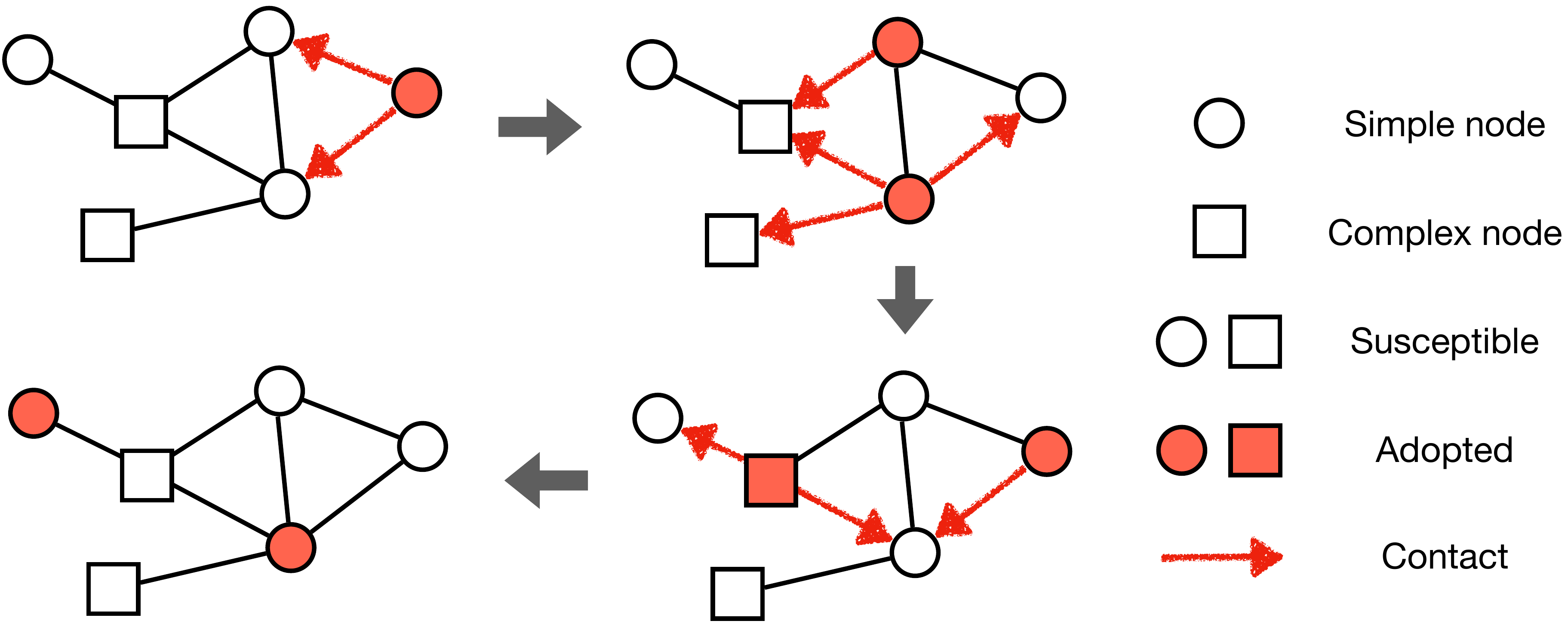}
\caption{
An example of our model with simple and complex nodes. Susceptible nodes (open symbols) 
are adopted when the number of contacts to the adopted is equal or larger than its 
assigned adoption threshold, either $n=1$ for simple nodes (circles) or $n=2$ for 
complex nodes (squares). Adopted nodes become susceptible in the next time step.
}
\label{fig:model}
\end{figure}

First, we consider a network with $N$ nodes that can be either susceptible or adopted. 
Thereafter, we assign adoption threshold $\theta_i$ for node $i$, which represents the number 
of contacts required to change their state from susceptible to adopted. When $\theta_i=1$, 
node $i$ is adopted after a single contact with an adopted neighbor at each time step 
according to simple contagion (simple nodes). When $\theta_i>1$, it corresponds to complex 
contagion, indicating that multiple contacts are required for adoption at each time step
(complex nodes). For simplicity, in this study, we assume that all complex nodes in 
a network have an adoption threshold of $\theta=n$. That is, the fraction $p$ of nodes 
are complex nodes with an adoption threshold of $n$, and the others $1-p$ are simple nodes.
The fraction $p$ of complex nodes represents the degree of the heterogeneity in adoption thresholds.

An example of the proposed model is presented in Fig.~\ref{fig:model}. The circles (squares) 
represent simple (complex) nodes and open (filled) symbols represent susceptible (adopted) 
nodes. Initially, the fraction $R_0$ of the nodes is adopted, and the others are susceptible. 
In our model, the dynamics is in discrete time steps. At each time step, each pair of 
connected nodes contacts each other with a contact probability $\lambda$. If the number 
of contacts with adopted neighbors is equal or larger than its adoption threshold, the node 
becomes adopted at the next time step. Contacts for adoption repeat for all nodes in a network. 
All adopted nodes become the susceptible state, which can be adopted again, like the 
recovery process in susceptible-infected-susceptible model. 
Note that an adopted node can remain adopted at the next time step if
the number of contacts to its adopted neighbors is equal or larger than its adoption threshold.
The adoption and recovery processes are repeated until a steady state is reached. 
In the steady state, the system is in either an absorbing or an active phase. 
In the absorbing phase, all nodes in a network 
are susceptible, and there is no more contagion dynamics. In contrast, a finite fraction of 
nodes remain in the adopted state in the active phase, which is the source of active dynamics. 
To assess the steady-state behavior, we measure the fraction $R$ of adopted nodes. While $R=0$ 
for the absorbing phase, $R$ is a nonzero value for the active phase.

The followings are the three important parameters of the model: 
contact probability $\lambda$, fraction $p$ of the complex nodes, and adoption 
threshold $n$ of the complex nodes. The contact probability $\lambda$ represents 
how often each connected pair interacts with each other, the fraction of complex nodes 
represents the degree of heterogeneous adoptability, and the adoption threshold $n$ 
represents the stubbornness of the complex nodes to adopt. We measure the fraction $R$ 
of the adopted nodes in a steady state based on three parameters, $\lambda$, $p$, and $n$.

\section{Analytical approach}

To predict the final fraction of adopted nodes, we derive heterogeneous mean-field 
equations, assuming a locally tree-like network in the limit $N \rightarrow \infty$. 
Because there are two classes of nodes in a network, we consider two types of node 
degrees: the number of links connected to the simple $k_s$ and complex $k_c$ neighbors. 
Additionally, we define $s_{k_s,k_c}^{t}$ and $c_{k_s,k_c}^{t}$ as the fraction of the
adopted
simple and complex nodes with degrees $k_s$ and $k_c$, respectively, at time $t$. 
We define the probability that a randomly chosen node pointing to a simple (complex)
node is an adopted one as $\phi_s^t$ ($\phi_c^t$).
For a random network without any degree-degree 
correlations, the probabilities $\phi_s^t$ and $\phi_c^t$ can be expressed as follows:
\begin{align}
\phi_s^t &= \sum_{k_s,k_c} \frac{k_s P(k_s,k_c)}{\langle k_s \rangle} s_{k_s,k_c}^t, \label{eq:phis} \\
\phi_c^t &= \sum_{k_s,k_c} \frac{k_c P(k_s,k_c)}{\langle k_c \rangle} c_{k_s,k_c}^t, \label{eq:phic}
\end{align}
where $P(k_s,k_c)$ is the distribution of the joint degree for $k_s$ and $k_c$ of a network.

The fraction of adopted simple nodes $s_{k_s,k_c}^{t+1}$ can be 
calculated according to the following equations \cite{gleeson2007,min2018}
\begin{align}
s_{k_s,k_c}^{t+1} &= 1- (1- \lambda \phi_s^{t})^{k_s} (1- \lambda \phi_c^{t})^{k_c}. 
\label{eq:s}
\end{align}
The terms $(1-\lambda \phi_s^{t})^{k_s}$ and $(1-\lambda \phi_s^{t})^{k_s}$ 
stand for the probability that a node does not contact with adopted simple and 
adopted complex neighbors, respectively. Similarly, the fraction of the adopted 
complex nodes can be obtained by 
\begin{align}
c_{k_s,k_c}^{t+1} &= 1- \sum_{\mu=0}^{n-1} \sum_{\mu_s=0}^{\mu} \binom{k_s}{\mu_s} 
	(\lambda \phi_s^{t})^{\mu_s} (1- \lambda  \phi_s^{t})^{k_s-\mu_s}  \nonumber \\
	\times& \binom{k_c}{\mu-\mu_s} ( \lambda \phi_c^{t})^{\mu-\mu_s} (1- \lambda \phi_c^{t})^{k_c-(\mu-\mu_s)}. 
\label{eq:c}
\end{align}
The term $\sum_\mu \cdots$ represents the probability that the number $\mu$ of contacts to adopted
neighbors is less than the adoption threshold $n$.
Note that if adopted nodes are again adopted by their neighbors, they
remain adopted at the next time step as Eqs.~\ref{eq:s} and \ref{eq:c} imply.

We can derive the following self-consistency equations in the limit $t \rightarrow \infty$
combining Eqs.~\ref{eq:phis}-\ref{eq:c}:
\begin{align}
\phi_s^\infty  &= 1- \sum_{k_s,k_c} \frac{k_s P(k_s,k_c)}{\langle k_s \rangle} 
(1- \lambda \phi_s^{\infty})^{k_s} (1- \lambda \phi_c^{\infty})^{k_c}, \\
\phi_c^\infty  &= 1- \sum_{k_s,k_c} \frac{k_c P(k_s,k_c)}{\langle k_c \rangle} 
\sum_{\mu=0}^{n-1} \sum_{\mu_s=0}^{\mu} \binom{k_s}{\mu_s} \binom{k_c}{\mu-\mu_s}  \\
	\times (\lambda &\phi_s^{\infty})^{\mu_s} (1- \lambda \phi_s^{\infty})^{k_s-\mu_s}  
	( \lambda \phi_c^{\infty})^{\mu-\mu_s} (1- \lambda \phi_c^{\infty})^{k_c-(\mu-\mu_s)}. \nonumber
\end{align}
We can obtain $\phi_s^{\infty}$ and $\phi_c^{\infty}$ by solving the coupled equations 
iteratively from the initial values of $\phi_s$ and $\phi_c$. Putting them into 
Eqs.~\ref{eq:s}-\ref{eq:c}, the fraction of the adopted simple and complex nodes
in the steady state, $s_{k_s,k_c}^{\infty}$, and $c_{k_s,k_c}^{\infty}$, can be
obtained. Finally, the average fraction $R$ of adopted nodes in the steady state 
with the fraction $p$ of complex nodes can be obtained by 
\begin{align}
R= \sum_{k_s,k_c} P(k_s,k_c) \left[ p c_{k_s,k_c}^{\infty} + (1-p) s_{k_s,k_c}^{\infty} \right].
\label{eq:r}
\end{align}

The fraction $R$ of the adopted nodes can be calculated from the fixed points in Eqs.~\ref{eq:phis} 
and \ref{eq:phic}. The trivial solution of $R$ with $\phi_s^\infty = \phi_c^\infty=0$ 
corresponds to an absorbing phase where $R=0$. An active phase with $R>0$ can appear when 
$\phi_c$ or $\phi_c$ is nonzero. 
The largest eigenvalue, $\Lambda$, of the Jacobian 
matrix $\mathcal{J}$ in Eqs.~\ref{eq:phis} and \ref{eq:phic} 
identifies the stability of a fixed point 
$\vec{\phi}=(\phi_s^*, \phi_c^*)$ by applying a linear stability analysis.
Specifically, the fixed point $(\phi_s^*, \phi_c^*)$ is stable when $\Lambda(\vec{\phi})>1$. 
Note that multiple stable fixed points of $\vec{\phi}$ can appear for a certain set of 
parameters.

A simple condition of the transition between the absorbing and active phases 
is that $\Lambda$ at $(\phi_s,\phi_c)=(0,0)$ is equal to unity. 
Since the Jacobian matrix $\mathcal{J}$ at $(\phi_s,\phi_c)=(0,0)$ is given by
\begin{align}
\mathcal{J} =
	\begin{pmatrix} 
		\lambda \frac{\langle k_s^2 \rangle}{\langle k_s \rangle} & \lambda \frac{\langle k_s k_c \rangle}{\langle k_s \rangle}  \\
		0 & 0 
	\end{pmatrix},
\end{align} this condition leads to the transition point for a given $p$
\begin{align}
\lambda_1=\frac{\langle k_s \rangle}{\langle k_s^2 \rangle}.
\end{align}
If $\lambda> \lambda_1$, the system reaches the active phase ($R>0$) initiated by adopted simple 
nodes. When all the nodes are simple nodes, we can recover the epidemic threshold for the 
susceptible-infected-susceptible model~\cite{pastor2001,pastor2015}.
There can be transition points $\lambda_{2}$ other than $(\phi_s,\phi_c)=(0,0)$. 
Unfortunately, we cannot express $\lambda_2$ in a closed form expression but the transitions 
can be identified by numerical integrations with the condition $\Lambda(\vec{\phi})=1$ for the non-zero 
solution of $\vec{\phi}=(\phi_s^*, \phi_c^*)$. It is not feasible to obtain $\lambda_2$ 
in a closed form because the non-zero fixed points $\vec{\phi}=(\phi_s^*, \phi_c^*)$ 
cannot be simply reduced into an analytical expression.
Also note that these transitions at $\lambda_2$ cause the abrupt change of $R$ 
between low $R$ and high $R$ rather than the transition between the absorbing and active phases.

\section{Results}

\subsection{Phase diagram on Erd\"os-R\'enyi graphs}

\begin{figure}
\includegraphics[width=\linewidth]{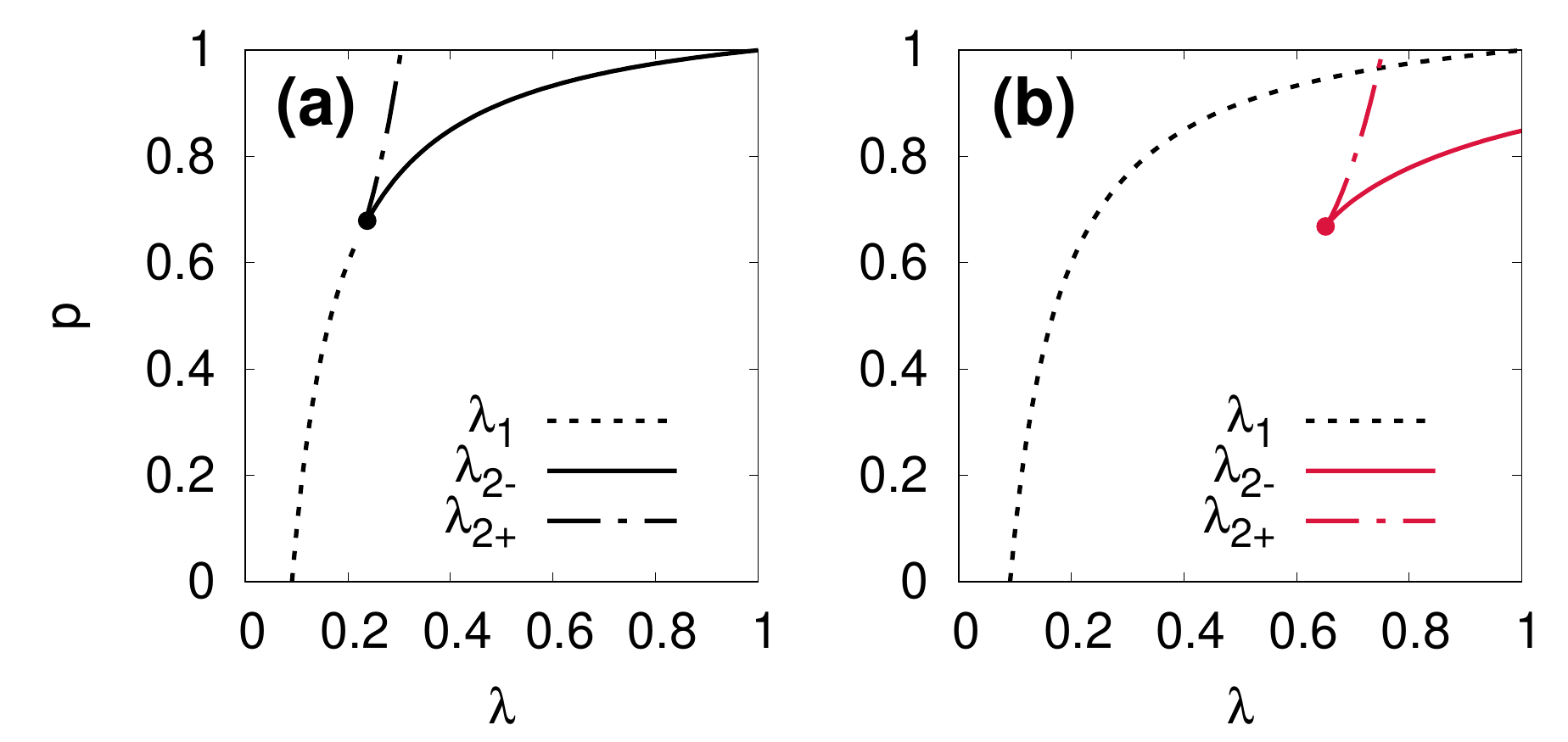}
\caption{
Phase diagram of a heterogeneous contagion model with (a) $n=2$ and (b) $n=5$ for ER
networks with $z=10$. Continuous transition lines $\lambda_1$ are indicated by dashed 
lines. Discontinuous transitions are indicated by dot-dashed lines $\lambda_{2+}$ 
and solid lines $\lambda_{2-}$. Critical points where the discontinuous jump 
disappears are indicated by filled circles.
}
\label{fig:phase}
\end{figure}

We consider the contagion model with heterogeneous adoption thresholds
on Erd\"os-R\'enyi (ER) networks. Degree distribution of the ER graphs in the 
thermodynamic limit $N \rightarrow \infty$ is approximately given by a Poisson 
distribution $P(k)=e^{-z}z^k/k!$, where $z$ denotes the average degree. We assume 
that simple and complex nodes are distributed randomly on a network. 
Thereafter, $P(k_s,k_c)$ can be decomposed into the product of two Poisson distributions, 
$P(k_s)$ and $P(k_c)$, with average degrees of $(1-p)z$ and $pz$, respectively. 
With the setting, the transition $\lambda_1$ is located at 
\begin{align}
\lambda_1=\frac{1}{(1-p)z+1}.
\label{eq:p1}
\end{align}

We examine the phase diagram with $n=2$ on ER networks with $z=10$ [Fig.~\ref{fig:phase}(a)]. 
First, we observe a continuous (dashed) transition line, as predicted by Eq.~\ref{eq:p1}. 
The transition becomes discontinuous, denoted by $\lambda_{2-}$ (solid), when the 
fraction $p$ of the complex nodes exceeds the critical point $(\lambda_c,p_c)$. In addition when 
$p>p_c$, another discontinuous transition line appeared, denoted by $\lambda_{2+}$ 
(dot-dashed). Our theory predicts that these three transition lines meet at a single point, 
$(\lambda_c,p_c)$. When $p$ is less than the point, there is only a continuous transition between 
the absorbing and active phases like a typical absorbing phase transition. However, there are 
two transition lines, $\lambda_{2-}$ and $\lambda_{2+}$ when $p$ is larger than $p_c$.
When the contagion dynamics begins from a tiny fraction of adopted seeds, that is 
$R_0 \approx 0.01$, the theory predicts the location of the transition at $\lambda_{2-}$.
However, when the fraction initially adopted nodes is sufficiently high, that is 
$R_0 \approx 0.99$, the transition takes place on $\lambda_{2+}$.

A phase diagram with $n=5$ is shown in Fig.~\ref{fig:phase}(b) as a typical example of $n>2$. 
We find qualitatively similar phase diagram for all $n>2$. We observe a continuous phase 
transition line $\lambda_1$, as predicted by Eq.~\ref{eq:p1}. In addition, we observe two 
additional discontinuous transition lines, $\lambda_{2+}$ (dot-dashed) and $\lambda_{2-}$ 
(solid), when $p$ is larger than the critical point $p_c$, denoted by a filled circle. 
Note that the discontinuous transition line $\lambda_{2-}$ is separated from $\lambda_1$. 
When $p<p_c$, a single transition point $\lambda_1$ is observed, meaning that the systems 
change from absorbing to active phase at $\lambda_1$. When $p>p_c$, a continuous transition 
line $\lambda_1$ and two discontinuous transitions lines, $\lambda_{2-}$ and $\lambda_{2+}$, 
are observed. Depending on the initial seed fraction, $R_0$, a discontinuous transition is 
realized at different points, either $\lambda_{2-}$ or $\lambda_{2+}$. Specifically, when 
dynamics starts with low $R_0$, the fraction $R$ of the adopted nodes abruptly changes at 
$\lambda_{2-}$. However when contagion processes starts with high $R_0$, the transition 
appears at $\lambda_{2+}$ and not $\lambda_{2-}$.

We conduct Monte-Carlo simulations of the contagion dynamics on ER networks with $N=10^4$ 
and $z=10$ for $n=5$ with various $p$ values, averaged over $10^4$ independent runs
in order to verify our theory. 
We find that the system reaches a steady state where the fraction $R$ of adopted nodes 
does not change significantly over time, after the initial transient period.
We measure the fraction of adopted nodes $R$ at the steady state. 
In this example, the dynamics starts with a few fractions of adopted seeds such as $R_0 \approx 0.01$. 
As shown in Fig.~\ref{fig:double}(a), 
numerical results (symbols) are consistent with the theory (lines). 
As the fraction $p$ of the complex nodes increases, the transition point $\lambda_1$ delays 
as predicted by theory. In addition, when $p>p_c$ where $p_c = 0.668$ for $n=5$,
an additional transition can appear at $\lambda_{2-}$. Note that $\lambda_{2+}$ is 
inaccessible with the choice of initial condition.

\subsection{Double transitions}

\begin{figure}
\includegraphics[width=\linewidth]{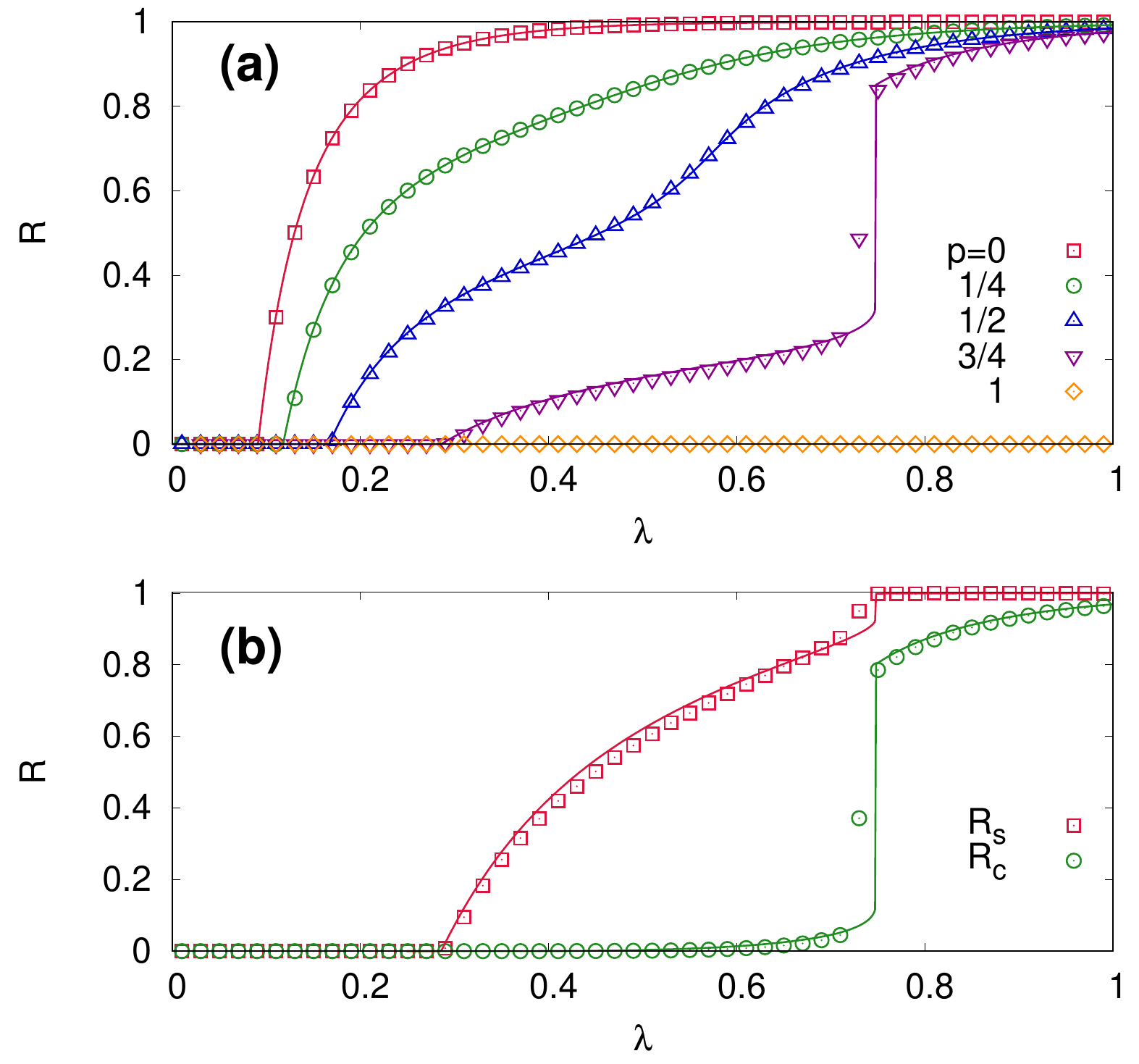}
\caption{
(a) Fraction $R$ of the adopted nodes as a function of $\lambda$
for different $p$ with $n=5$, $R_0 \approx 0.01$, on ER networks with $N=10^4$, 
and $z=10$, averaged over $10^4$ independent runs. 
(b) Fraction of adopted simple nodes $R_s$ (squares) and complex nodes $R_c$ (circles) 
as a function of $\lambda$ with $p=3/4$. In the intermediate phase, $R_s$ increases
gradually whereas most complex nodes remain susceptible (low $R_c$).
Numerical simulations (symbols) and theoretical calculation (lines) are shown together. 
}
\label{fig:double}
\end{figure}

Let us examine the contagion dynamics with $n=5$ and $p=3/4$, which is $p>p_c$. 
As shown in Fig.~\ref{fig:double}(a), we have two consecutive transitions, $\lambda_1$ 
and $\lambda_{2-}$, which are called a double transition \cite{colomer2014,allard2017,min2018}. 
The transition between the absorbing and active phases takes place at $\lambda_1$. 
In addition, the fraction $R$ of the adopted nodes that are already non-zero suddenly increases 
at $\lambda_{2-}$ with a discontinuous jump in $R$. There is an intermediate phase between the 
two transitions, where $R$ is non-zero but still low. In this model, we show
that the double phase transitions can naturally appear because of the heterogeneity of 
the adoption thresholds.
For social contagions, it is often more realistic to use the threshold 
as the relative fraction of contacts to adopted nodes out of the total number of neighbors rather 
than the absolute number of contacts to the adopted \cite{granovetter1978,watts2002}. 
We observed that the double phase transitions also appear when we use the threshold 
given by the fraction of adopted nodes (see appendix).

We measure the fraction of adopted nodes with simple contagion $R_s$ and complex contagion 
$R_c$ with $p=3/4$ to demonstrate the mechanism of the double phase transition 
[Fig.~\ref{fig:double}(b)]. Here, $R_s$ ($R_c$) represents the fraction of the adopted simple 
(complex) nodes for all simple (complex) nodes. For instance, $R_s=0$ when all the simple nodes 
are susceptible, and $R_s=1$ when all the simple nodes are adopted. In an absorbing 
phase $\lambda<\lambda_1$, all nodes are susceptible regardless of being simple or complex nodes,
leading to $R=0$. At $\lambda_1$, simple nodes start to be adopted. However, most complex nodes 
remain susceptible (low $R_c$) until $\lambda<\lambda_{2-}$. Therefore, simple nodes are adopted 
whereas most complex nodes are susceptible in the intermediate phase, $\lambda_1 < \lambda < \lambda_{2-}$. 
At the second transition $\lambda_{2-}$, complex nodes collectively change their state from 
susceptible to adoption, leading to a discontinuous jump in $R_c$. As a result, when 
$\lambda>\lambda_{2-}$, most nodes, either simple or complex, are adopted. Therefore,
both $R_s$ and $R_c$ show a high value with a discontinuous jump at $\lambda_{2-}$.

\subsection{Hysteresis curves}

\begin{figure}
\includegraphics[width=\linewidth]{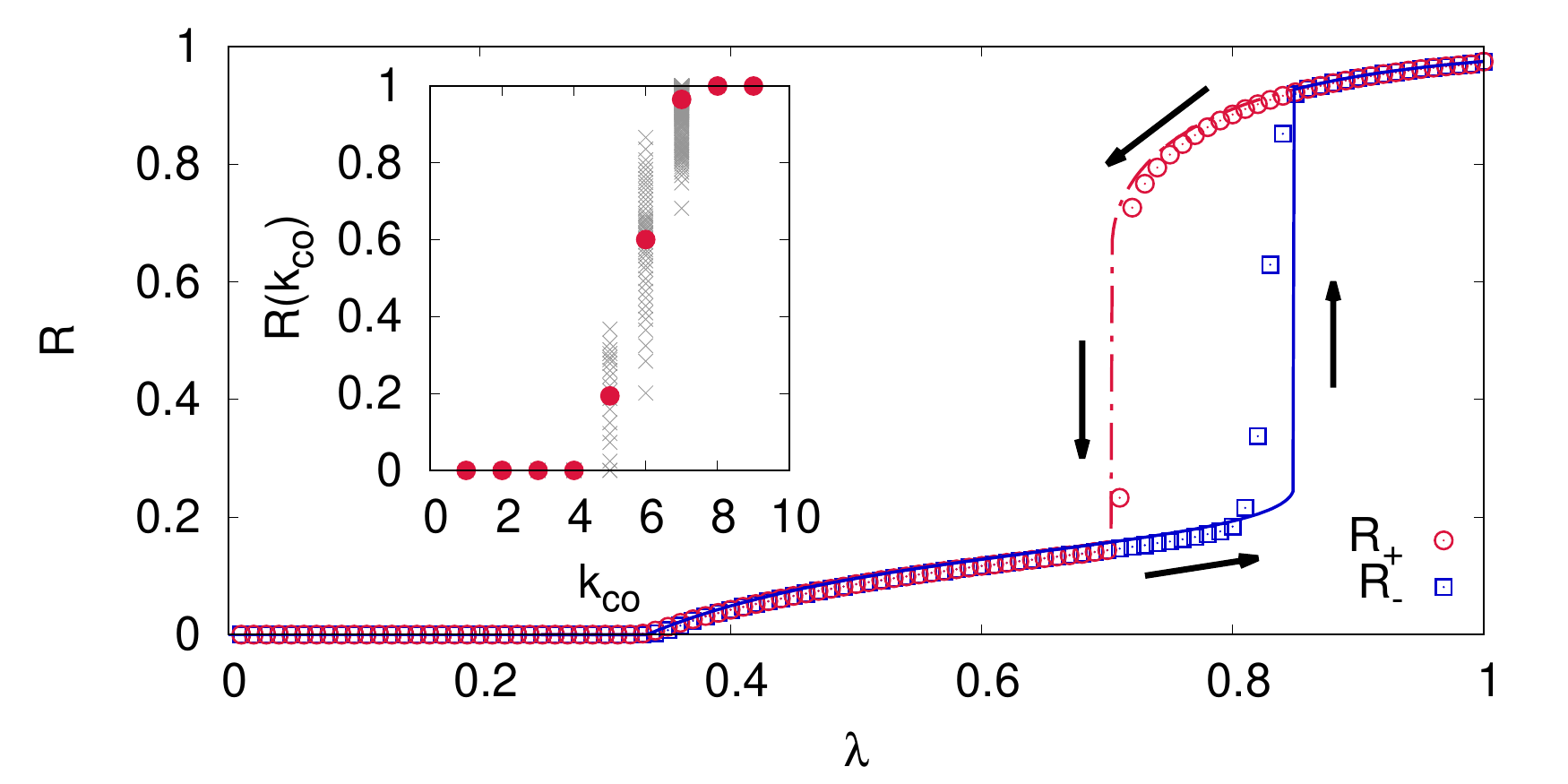}.
\caption{
The fraction of adopted nodes $R$ as a function of $\lambda$ for ER 
graphs with $N=10^4$, $z=10$, $n=5$, and $p=0.8$. We find the bistable region 
where two stable $R$, $R_+$ and $R_-$, exist between the two transitions,
$\lambda_{2+}$ and $\lambda_{2-}$. Numerical simulations (symbols)
and theoretical calculation (lines) are shown. (inset) The fraction of the 
adopted complex nodes at $\lambda=0.8$, following $R_{+}$ curve averaged over 
nodes with the same $k$-core index $k_{co}$ (filled circles). Gray symbols 
represent the probability that a node with $k_{co}$ is adopted and obtained 
for each node averaged in time.
}
\label{fig:hys}
\end{figure}

Moreover, we show the implication in two discontinuous transition lines, $\lambda_{2-}$
and $\lambda_{2+}$, above $p_c$ [see Fig.~\ref{fig:phase}(b)]. There are two different $R$ 
in a steady state between two transition lines, meaning that $\lambda_{2+}<\lambda<\lambda_{2-}$. 
As shown in Fig.~\ref{fig:hys}, there is a bistable region with two stable $R$, for instance, 
$\lambda \approx 0.8$. We defined two loci of stable $R$ as $R_+$ (circles) and $R_-$ (squares). 
While we can obtain the upper locus, $R_+$, when contagion dynamics starts with a high $R_0$, 
we can arrive at the lower locus, $R_-$, when $R_0$ is low. Depending on the initial fraction 
of the seed nodes, the locations of the transition point where a discontinuous jump appears 
is also different, either $\lambda_{2+}$ for $R_+$ or $\lambda_{2-}$ for $R_-$.
Here we use ER networks with $N=10^4$, $z=10$ for $n=5$ and $p=0.8$ over $10^4$ independent 
runs in our numerical simulations.
And, the dynamics starts with a fractions of adopted seeds $R_0 \approx 0.01$ for $R_-$
and $R_0 \approx 0.99$ for $R_+$.

Because of bistability, there can be hysteretic behavior in the fraction $R$ of 
the adopted nodes by varying $\lambda$. If the transmission probability increases
from low $\lambda$, the system sustains the intermediate phase where simple nodes 
are adopted but most complex nodes remain susceptible until $\lambda_{2-}$. At 
$\lambda_{2-}$, $R$ suddenly increases because of the adoption of complex nodes. 
But when the system reaches high $R$ phase, the phase maintains for 
$\lambda_{2+}<\lambda<\lambda_{2-}$. 
To regain the low $R$ phase, $\lambda$ must be less than $\lambda_{2+}$. 

We search for the origin of hysteretic behavior based on the topology of a network.
We first assign the $k$-core (also known as the $k$-shell) index for each node to assess 
the coreness \cite{dorogovtsev2006}. Here, $k$-core represents a subset of nodes 
formed by iterative removal of all nodes with degree less than $k$. 
That is, the $k$-core is a maximal set of nodes where all nodes have at least $k$ degrees
within the set. The procedure to assign $k$-core index is as follows:
i) First, assign $k$-core index to each node as $k_{co}=1$.
ii) Remove all nodes with remaining degree $k \le k_{co}$, iteratively.
iii) Increase $k$-core index by one to all remaining nodes.
iv) Repeat ii) and iii) until all nodes are removed.
Thereafter, a unique value of $k_{co}$ is assigned for each node.
The $k$-core index assigned represents the coreness of each node in the topology 
of a network.

We measure the average fraction $R(k_{co})$ of the adopted complex nodes as a function 
of the $k$-core index $k_{co}$ in the bistable region, that is, $p=0.8$. The inset 
of Fig.~\ref{fig:hys} shows $R(k_{co})$ in the upper locus, $R_+$. We find that 
nodes with a high $k$-core index can be in the adopted state for $R_+$, implying 
that the complex nodes in the core of a network can collectively sustain the active state. 
Additionally, once high $k$-core nodes in $R_+$ become susceptible, $R$ remains 
at a low value until a group of complex nodes 
collectively change their state to be adopted at $\lambda_{2-}$. 

\section{Discussion}

Here, we study a model of contagion dynamics with heterogeneous adoption 
thresholds for different agents. In addition, we incorporate a recovery process from adopted
to susceptible state. We find a double transition, which is a continuous transition
from the absorbing to active phases and a subsequent discontinuous jump in the 
fraction of the adopted nodes. The double transition occurs with an intermediate phase 
in which simple nodes are adopted but complex nodes remain susceptible. Moreover, we 
find hysteresis in the fraction of the adopted nodes with respect to the contact 
probability. 
Our study sheds light on some fundamental aspects of the mechanism
of hysteresis in contagion processes and multiple phase transitions in complex networks.
For more realistic modeling it would be also interesting to investigate 
these factors that we has not covered in this study: the influence of heterogeneity 
in degree distributions such as scale-free networks and/or the existence 
of degree-degree correlations, to name a few.


\begin{acknowledgments} 
This work was supported by the National Research Foundation of Korea (NRF) grant funded 
by the Korean Government (MSIT) (2018R1C1B5044202 and 2020R1I1A3068803).
\end{acknowledgments}

\appendix*

\section{Threshold Model}

It is often more realistic to define the threshold as the relative fraction of adopted 
nodes out of the total number of neighbors rather than the number of adopted nodes, in 
particular for social contagions \cite{granovetter1978,watts2002}. 
In order to validate the robustness of our main finding, we check the
case with the threshold defined by the fraction of contacts to adopted nodes.
To be specific, for complex node $i$, when the fraction 
of contacts to adopted nodes is equal or larger than a certain threshold $\eta$,
node $i$ becomes adopted. We assume as previous that the fraction $p$ of nodes are complex 
nodes with an adoption threshold of $\eta$ for the fraction of adopted contacts, 
and the others $1-p$ are simple nodes.

\begin{figure}
\includegraphics[width=\linewidth]{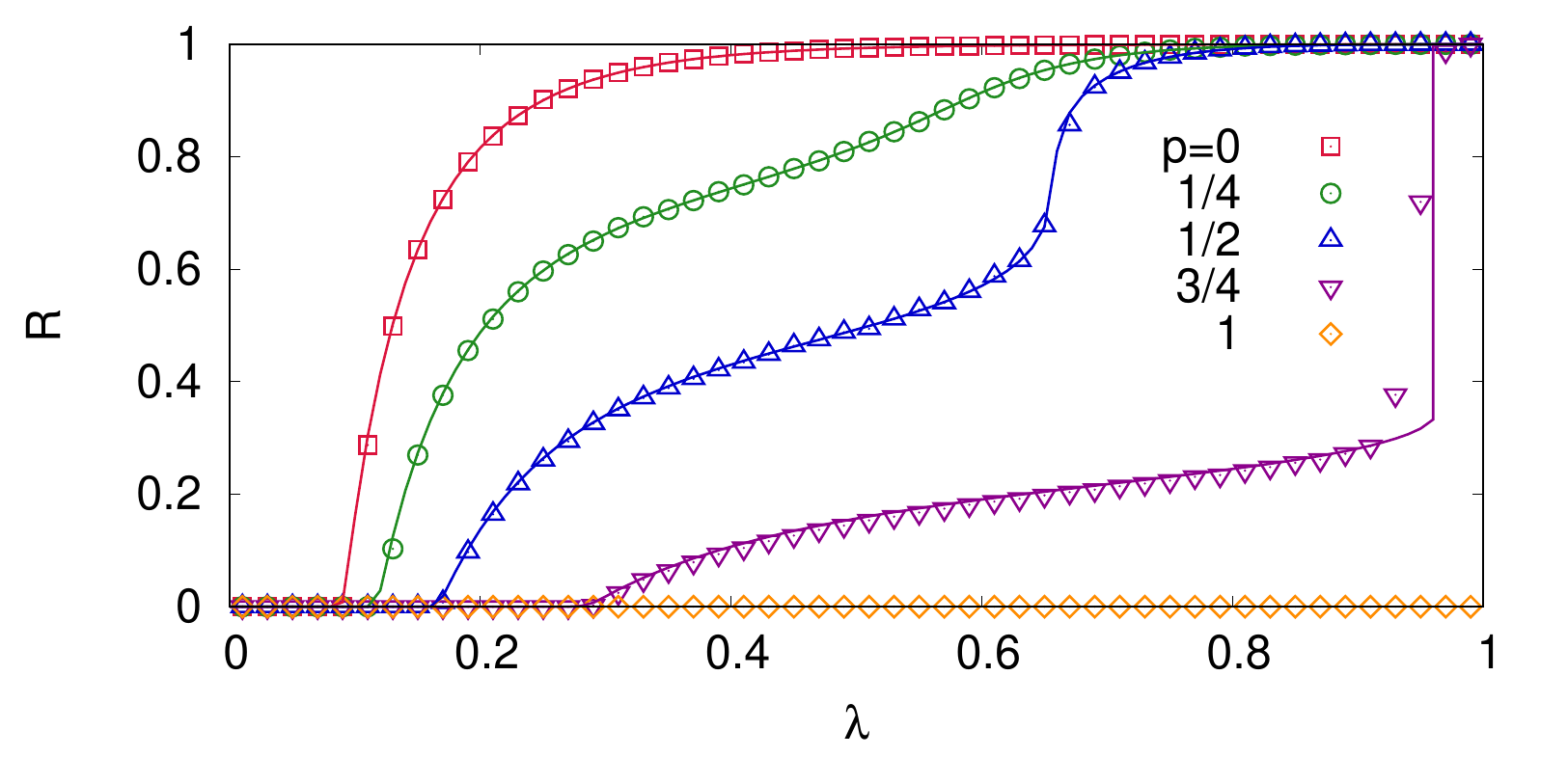}
\caption{
Fraction $R$ of the adopted nodes as a function of $\lambda$
for different $p$ with $\eta=1/2$, $R_0 \approx 0.01$, on ER networks with $N=10^4$
and $z=10$, averaged over $10^2$ independent runs. 
}
\label{fig:appendix}
\end{figure}

For the case, we can derive the similar equations with Eqs.~\ref{eq:phis}-\ref{eq:c}
with the following modification of $c_{k_s,k_c}$:
\begin{align}
c_{k_s,k_c}^{t+1} &= 1- \sum_{\mu=0}^{\mu/(k_s+k_c)<\eta} \sum_{\mu_s=0}^{\mu} \binom{k_s}{\mu_s} 
	(\lambda \phi_s^{t})^{\mu_s} (1- \lambda  \phi_s^{t})^{k_s-\mu_s}  \nonumber \\
	\times& \binom{k_c}{\mu-\mu_s} ( \lambda \phi_c^{t})^{\mu-\mu_s} (1- \lambda \phi_c^{t})^{k_c-(\mu-\mu_s)}. 
\label{eq:appen}
\end{align}
Then, by using Eq.~\ref{eq:r}, we can obtain the fraction of adopted nodes in the steady
state with the threshold given by the fraction. We also conduct numerical simulations 
of the dynamics on ER networks with $N=10^4$ and $z=10$ for $\eta=1/2$, $R_0 \approx 0.01$ 
with various $p$ values, averaged over $10^2$ independent runs. As shown in Fig.~\ref{fig:appendix},
numerical results (symbols) are well consistent with the theory (lines). 
In addition, the double phase transitions are observed for the threshold 
given by the fraction of adopted nodes.

\end{document}